\documentclass[aps,amsfonts,amsmath,prd,preprint,nofootinbib,tightenlines,showpacs]{revtex4}
\usepackage{epsf}

\def\be{\begin{equation}}
\def\ee{\end{equation}}
\def\bea{\begin{eqnarray}}
\def\eea{\end{eqnarray}}
\def\bml{\begin{subequations}}
\def\blea{\bml\begin{eqnarray}}
\def\elea{\end{eqnarray}\end{subequations}}

\begin{document}

\title{Null energy conditions outside a background potential}

\author{Delia Schwartz Perlov}
\email{Delia.Perlov@tufts.edu}
\author{Ken D.\ Olum}
\email{kdo@cosmos.phy.tufts.edu}
\affiliation{Institute of Cosmology,
Department of Physics and Astronomy,
Tufts University, Medford, MA  02155}

\begin{abstract}
We study the ``Casimir'' energy of a minimally coupled, real, massless
scalar field outside a spherically symmetric background potential.  We
obtain a general expression for the null energy condition in $d$
dimensions and explicit expressions for a perfectly reflecting
spherical boundary in 3+1 and 2+1 dimensions.  In these cases, the
null energy condition is always violated for radial motion and
obeyed for azimuthal motion.  Nevertheless, the averaged null energy
condition is always obeyed.
\end{abstract}


\maketitle

\section{Introduction}

There are several energy conditions in General Relativity which try to
make a precise statement of the common sense notion that matter has
positive energy density.  The weak energy condition (WEC) states that
no observer sees negative energy density, in other words that
\be\label{eqn:WEC}
T_{\mu\nu}V^\mu V^\nu \geq {0}\,,
\ee
where $T_{\mu\nu}$ is the stress-energy tensor and $V^\mu$ is any
timelike vector.  If instead we consider null $V^\mu$ in Eq.\
(\ref{eqn:WEC}), we get the null energy condition (NEC).

The energy density between parallel conducting plates in the
traditional Casimir effect \cite{casimir} is negative, and thus
provides one example of a system for which the WEC is violated.
The pressure in the Casimir system is negative in the direction
between the plates, so with $V^\mu$ pointing from one plate toward the
other, the NEC is also violated.  However, null vectors parallel
to the plates have $T_{\mu\nu}V^\mu V^\nu = {0}$ because the
positive pressure in that direction cancels the negative energy
density.

The averaged weak energy condition (AWEC) permits the violation of
the WEC locally, but insists that when one integrates
the WEC over a complete geodesic with tangent vector $V^\mu$, the result will be non-negative,
\be\label{eqn:AWEC}
\int_{-\infty}^\infty dx T_{\mu\nu}V^\mu V^\nu \geq {0}\,.
\ee
The averaged null energy condition (ANEC) is just Eq.\
(\ref{eqn:AWEC}) for null geodesics.

One can imagine an observer at rest between the plates in the Casimir
system, so that $V^\mu$ points only in the time direction.  For that
observer, the AWEC is violated.  However, the Casimir system does not
violate the ANEC, because geodesics parallel to the plates obey the
NEC, while those that are not parallel eventually intersect the
plates and pick up a large positive contribution from the plate
material.

The ANEC is in a certain sense the weakest of these conditions.  If
the ANEC is violated then the NEC must be violated somewhere, and so the WEC
must also be violated, or else the NEC would hold by continuity.  For a
localized system in flat space, ANEC violation implies AWEC violation
by the same continuity argument, but otherwise
they are independent \cite{Roman}.  Nevertheless, the ANEC is
sufficient to rule out many exotic phenomena, such as traversable
wormholes \cite{wormholes}, superluminal travel\footnote{In this case,
the relevant integral is not over a complete geodesic, but over the
path to be traveled superluminally.} \cite{superluminal}, and closed
timelike curves \cite{Hawking}, and to prove singularity theorems
\cite{singularity}.

It is very important to understand under what conditions the
averaged null energy condition is violated, as Klinkhammer urged
in \cite{Klinkhammer}.  Klinkhammer found that for a \emph{free}
scalar field in Minkowski spacetime, the ANEC is satisfied.
However, he found ANEC violations for the free scalar field in a
flat cylindrical topology, which corresponds to closing Minkowski
space in one spatial direction.

We study here a Quantum Field Theory (QFT) problem of a minimally
coupled real massless scalar field outside a spherically symmetric
background potential (i.e., a field with a position-dependent
mass), in the calculational framework developed in
\cite{OlumGraham}.  In Sec.\ \ref{sec:nec} we obtain a general
expression for the null projection of the stress-energy tensor in
arbitrary dimension, followed by the results for 3+1 dimensions.
In Sec.\ \ref{sec:hard} we specialize to the case of a perfectly
reflecting sphere in 3+1 dimensions.  We find that the NEC is
violated for radial motion and obeyed for azimuthal motion, whilst
its validity for intermediate directions depends on the distance
from the sphere. We then calculate the ANEC by integrating over a
geodesic that passes outside the sphere. Although the NEC is
violated far from the sphere where the motion is primarily radial,
the points closer to the sphere dominate, and as a result the ANEC
is always obeyed.

In Sec.\ \ref{sec:2d} we briefly report our calculations of the NEC
and ANEC outside a perfectly reflecting circle in 2+1 dimensions.  The
results are very similar to those for 3+1 dimensions, and once again
we find that although the NEC may be violated, the ANEC is satisfied.


\section{Null energy condition for a scalar field}
\label{sec:nec}

\subsection{General case}
In this section we find a general expression for the NEC, for the
case of a minimally coupled, real scalar field outside
a potential.
The Lagrangian is
\be
{\cal L} = \frac{1}{2}\left[\partial_\mu\phi\partial^\mu\phi
-V(x)\phi^2)\right]
\ee
and the stress-energy tensor is
\be \label{eqn:stressenergytensor}
T_{\mu\nu}=\partial_\mu\phi\partial_\nu\phi+\frac{1}{2}\eta_{\mu\nu}\left[V(x)-\partial^\lambda\phi\partial_\lambda\phi\right]\,.
\ee
For a null vector,
$\eta_{\mu\nu}V^\mu V^\nu= 0$, so we have
\be
T_{\mu\nu}V^\mu V^\nu= \left(V^\alpha\partial_\alpha\phi \right)^2\,.
\ee
For a static system\footnote{Greek letters run over all indices and
Latin ones run over spacelike indices only.}, $T_{0i} = 0$.  If we further choose
coordinates in which $T_{ij}$ is diagonal, and $V = (1,{\bf v})$,
then
\be
T_{\mu\nu}V^\mu V^\nu = \dot\phi^2+ \sum_i
\left(v_i\partial_i\phi \right)^2\,.
\ee

\subsection{Spherical symmetry}

We are going to calculate the NEC for the scalar field outside a
potential that is spherically symmetric in $m$ spatial dimensions.
Using spherical coordinates and choosing $V$ to lie in the
equatorial plane, we find
\be \label{eqn:sphericalnec}  T_{\mu\nu}V^\mu V^\nu = \dot\phi^2+
\left(v_r\partial_r\phi \right)^2 +\left(v_{\varphi}\frac{1}{r}
\partial_{\varphi} \phi \right)^2
\ee
where $v_r$ is the component of the velocity in the radial direction
and $v_\varphi$ is the azimuthal component.  Since we are taking $V^t
= 1$, $v_\varphi^2 + v_r^2 = 1$.

Decomposing the quantum field $\phi$ in terms of modes gives
\bea
\label{eqn:scalarfield} \phi(r,\Omega,t) &=&
\sum_{\ell,\ell_z} \sqrt{\frac{2
\pi^{\frac{m}{2}}}{\Gamma\left(\frac{m}{2}\right)}} \\
 &&\times \int_0^\infty \frac{dk}{\sqrt{2 \pi \omega}} \left(
\psi^\ell_k(r)^* Y^m_{\ell\ell_z}(\Omega)^* e^{i\omega
t} a_{k}^{\ell\ell_z}{}^\dagger + \psi^\ell_k(r)
Y^m_{\ell\ell_z}(\Omega)
 e^{-i\omega t} a_{k}^{\ell\ell_z}\right) \nonumber
\eea
where $\omega = k$ since the field is massless, and the sum over
$\ell$ gives the partial wave expansion in the $m$ spatial dimensions.

The wavefunctions $\psi^\ell_k(r)$ are the eigenstates of the
time-independent radial Schr\"{o}dinger equation
\be\label{eqn:Schrodinger}
\left(-\frac{d^2}{dr^2} - \frac{m-1}{r}\frac{d}{dr} +
\frac{\ell(\ell+m-2)}{r^{2}} + V(r)\right) \psi^\ell_k(r) = k^2
\psi^\ell_k(r)\,.
\ee
In general, the solutions to Eq.\ (\ref{eqn:Schrodinger}) comprise both
bound and scattering states, but here we will consider only repulsive
potentials, so there are no bound states.

The wavefunctions and creation and annihilation operators are
normalized as follows:
\blea
\int Y^m_{\ell\ell_z}(\Omega)^\ast Y^m_{\ell'\ell'_z}(\Omega) \,
d\Omega &=& \delta_{\ell \ell'} \delta_{\ell_z \ell'_z}\\
\label{eqn:norm} \frac{2
\pi^{\frac{m}{2}}}{\Gamma\left(\frac{m}{2}\right)} \int_0^\infty
r^{m-1} \psi^\ell_k(r)^* \psi^{\ell'}_{k'}(r) \, dr &=& \pi
\delta(k-k') \\
{}[a_{k}^{\ell\ell_z}{}^\dagger, a_{k'}^{\ell'\ell_z'}{}^\dagger] &=&
[a_{k}^{\ell\ell_z}, a_{k'}^{\ell'\ell_z'}] = 0\\
{}[a_{k}^{\ell\ell_z}, a_{k'}^{\ell'\ell_z'}{}^\dagger] &=&
\delta(k-k') \delta_{\ell \ell'} \delta_{\ell_z \ell_z'}\,.
\elea

Using these expressions, the vacuum expectation of the time
derivative term in Eq.\ (\ref{eqn:sphericalnec}) is found
to be
\be
\label{eqn:timederiv} \frac{1}{2} \sum_\ell
D^m_\ell\int_0^\infty \frac{dk}{\pi}\omega \left(
|\psi^\ell_k(r)|^2 - |\psi^\ell_k{}^{(0)}(r)|^2 \right)\,.
\ee
where we have used dimensional regularization \cite{OlumGraham},
and have subtracted the free wavefunctions $\psi^\ell_k{}^{(0)}(r)$.
No other counterterms are necessary, because we are considering
only locations outside the potential.
The factor $D^m_\ell$ is the degeneracy in each partial wave
\cite{OlumGraham}:
\be
D^m_\ell = \frac{\Gamma(m+\ell-2)}{\Gamma(m-1)\Gamma(\ell+1)}(m+2\ell-2)
\ee
with $D^2_0=1$.

The vacuum expectation of the radial term of  Eq.\
(\ref{eqn:sphericalnec}) gives
\be
\label{eqn:radialterm} \frac{1}{2} v_r^2 \sum_\ell
D^m_\ell\int_0^\infty \frac{dk}{\pi}
\frac{1}{\omega}\left(|\partial_r\psi^\ell_k(r)|^2-|\partial_r\psi^{\ell(0)}_k(r)|^2\right)\,.
\ee
In the azimuthal term of  Eq.\ (\ref{eqn:sphericalnec}) we have to
differentiate the spherical harmonic, $Y^m_{\ell\ell_z}(\Omega)$.
In $2$+$1$ dimensions, $Y^m_{\ell\ell_z}(\Omega)\propto
e^{i\ell_z\varphi}$.  The same result holds for $3$+$1$ dimensions when we
consider the specific case of geodesics in the equatorial plane.
Thus, the vacuum expectation of the azimuthal term gives
\be
\label{eqn:azimuthterm} \frac{1}{2} v_\varphi^2 \sum_\ell
C^m_\ell\int_0^\infty \frac{dk}{\pi} \frac{1}{\omega r^2} \left(
|\psi^\ell_k(r)|^2 - |\psi^\ell_k{}^{(0)}(r)|^2 \right) \ee where
\be C^m_\ell =\sum\ell_z^2 =
\begin{cases}
2\ell^2 & m = 2\cr \ell (\ell+1) (2\ell+1)/2 & m = 3\,.
\end{cases}
\ee
Adding Eqs.\ (\ref{eqn:timederiv}), (\ref{eqn:radialterm}) and
(\ref{eqn:azimuthterm}), we find
\bea
\label{eqn:vacexpnec}  T_{\mu\nu}V^\mu V^\nu
= \frac{1}{2}\sum_\ell \int_0^\infty \frac{dk}{\pi} \bigg[
&&\left( D^m_\ell\omega + C^m_\ell \frac{v_\varphi^2}{\omega
r^2}\right) \left( |\psi^\ell_k(r)|^2 - |\psi^\ell_k{}^{(0)}(r)|^2
\right)\cr &&+D^m_\ell \frac{v_r^2}{\omega}
\left(|\partial_r\psi^\ell_k(r)|^2-|\partial_r\psi^{\ell(0)}_k(r)|^2\right)
\bigg]\,.
\eea

Outside any spherically symmetric potential, the wave functions
are given by
\be \label{eqn:wavefn}
\psi^\ell_k(r)= \sqrt{N_m(r)k} \left[{e^{2
i \delta_\ell}}H^{(1)}_\nu(kr) + H^{(2)}_\nu(kr)\right]
\ee
where $\delta_\ell$ is the scattering phase shift in the quantum
mechanical problem with the same potential,
$\nu =m/2-1+\ell$, and the normalization factor
\be
N_m(r) = \frac{\Gamma(m/2)}{8\pi^{m/2-1} r^{m-2}}\,.
\ee
The free wavefunctions $\psi^{\ell(0)}_k(r)$ are given by  Eq.\
(\ref{eqn:wavefn}), with $\delta_\ell=0$.
Thus we find
\be \label{eqn:diffwavefn}
|\psi^\ell_k(r)|^2-|\psi^{\ell(0)}_k(r)|^2= N_m(r)k\left[
\left(e^{2i \delta_\ell}-1\right)H^{(1)}_\nu(kr)^2 +\left(e^{-2i
\delta_\ell}-1\right)H^{(2)}_\nu(kr)^2\right]\,.
\ee

The radial derivative in Eq.\ (\ref{eqn:vacexpnec}) is
\be
\partial_r\psi^\ell_k(r)= \sqrt{N_m(r)k^3}
\left[{e^{2i\delta_\ell}} \tilde H^{(1)}_\nu(kr) + \tilde
H^{(2)}_\nu(kr)\right]
\ee
where
\be
\label{eqn:hankeltilde} \tilde
H^{(1)}_\nu(z)=\frac{1-m/2}{z}H^{(1)}_\nu(z) +H^{(1)}_\nu{}'(z)
=\frac{\ell}{z}H^{(1)}_\nu(z) -H^{(1)}_{\nu+1}(z)\,.
\ee
and likewise for $H^{(2)}_\nu$.
Therefore
\be
\label{eqn:diffderivwavefn}
|\partial_r\psi^{\ell}_k(r)|^2-|\partial_r\psi^{\ell(0)}_k(r)|^2=N_m(r)k^3
\left[\left(e^{2i \delta_\ell}-1\right) \tilde
H^{(1)}_\nu(kr)^2+\left(e^{-2i \delta_\ell}-1\right) \tilde
H^{(2)}_\nu(kr)^2\right]\,.
\ee

We show in the appendix that the second term of Eq.\
(\ref{eqn:diffwavefn}) is just the first term with the replacement
$k\rightarrow-k+i\epsilon$ and likewise for Eq.\
(\ref{eqn:diffderivwavefn}).  Thus we can drop the second terms and
extend the range of integration over $k$ to $-\infty$, with the
understanding that $k$ is to be taken above any branch cut on the
negative real axis.  Then
\bea
T_{\mu\nu}V^\mu V^\nu = \frac{1}{2}\sum_\ell
\int_{-\infty}^\infty & &\frac{dk}{\pi} \left(e^{2i
\delta_\ell}-1\right) N_m(r)k\nonumber\\
& &\times \bigg[\left( D^m_\ell\omega+
C^m_\ell \frac{v_\varphi^2}{\omega r^2}\right)
H^{(1)}_\nu(kr)^2+D^m_\ell\frac{v_r^2 k^2}{\omega} \tilde
H^{(1)}_\nu(kr)^2 \bigg]\,.
\eea
Following the methods used in \cite{OlumGraham}, we now convert
this expression to a contour integral which we close in the upper
half plane.  In general, $\delta_l(k)$ will not be well behaved off the
real axis, but in the present case of a potential with compact
support there will be no difficulty.
The only contribution to the integral comes
from the branch cut along the imaginary $k$ axis.  To the right
$\omega = \sqrt{k^2} = k$, but to the left $\omega = -k$, so with $k = i\kappa$, we obtain
\bea
 T_{\mu\nu}V^\mu V^\nu =
-\frac{N_m(r)}{\pi}\sum_\ell\int_0^\infty d\kappa& &\,i \left(e^{2i
\delta_\ell(i\kappa)}-1\right)\kappa ^2  \\
&&\times \left[ \left( D^m_\ell -C^m_\ell \frac{v_\varphi^2}{\kappa^2
r^2}\right) H^{(1)}_\nu(i\kappa r)^2 +D^m_\ell v_r^2 \tilde
H^{(1)}_\nu(i\kappa r)^2\right]\,.\nonumber
\eea
Using Eq.\ (\ref{eqn:hankeltilde}) and $H^{(1)}_\nu(ix) =
(2/\pi)i^{- (\nu+1)} K_\nu(x)$, we get
\bea
\label{eqn:gennec} T_{\mu\nu}V^\mu V^\nu =
&&-\frac{4N_m(r)}{\pi^3}\sum_\ell i^{-{2\nu-1}}\int_0^\infty
d\kappa\, \left(e^{2i \delta_\ell(i\kappa)}-1\right)\kappa^2
\left[ \left( D^m_\ell -C^m_\ell \frac{v_\varphi^2}{\kappa^2
r^2}\right)
K_\nu(\kappa r)^2\right.\\
& &\hspace{2.9in}\left.-D^m_\ell v_r^2
\left(\frac{\ell}{\kappa r}K_\nu(\kappa r)-K_{\nu+1}(\kappa r)\right)^2\right]
\nonumber
\eea
for the general case of a minimally coupled, massless, scalar
field outside a spherically symmetric potential.

For $m = 3$, we find $\nu = l+1/2$, $N_m (r) = 1/(16r)$, $D^m_\ell
= 2\ell+1$, $C^m_\ell = \ell (\ell+1) (2\ell+1)/3$, and
\bea
\label{eqn:m3gennec} T_{\mu\nu}V^\mu V^\nu
= \frac{1}{4\pi^3r}\sum_\ell (-1)^l \int_0^\infty & d\kappa&
\left(e^{2i \delta_\ell(i\kappa)}-1\right)\kappa^2\\
&\times& \left[\left(2\ell+1 -\frac{\ell (\ell+1)
(2\ell+1)v_\varphi^2}{2\kappa^2
r^2}\right) K_{\ell+1/2}(\kappa r)^2 \right. \nonumber \\
&& \left. - (2\ell+1) v_r^2 \left(\frac{\ell}{\kappa
r}K_{\ell+1/2}(\kappa r)-K_{\ell+3/2}(\kappa r)\right)^2\right]\,.\nonumber
\eea

In Sec.\ \ref{sec:2d} we will summarize the results for 2+1 dimensions,
which closely mimic our findings for 3+1 dimensions.


\section{Perfectly Reflecting Boundary Conditions}
\label{sec:hard}
\subsection{Phase shifts}

For a hard sphere of radius $a$, the scattering phase shift is
\be
{e^{2 i \delta_\ell}} = -
\frac{H^{(2)}_\nu(ka)}{H^{(1)}_\nu(ka)}
\ee
so that the wave function vanishes at $r=a$.  Thus
\be
{e^{2 i \delta_\ell}}-1 = - \frac{2J_\nu(ka)}{H^{(1)}_\nu(ka)}\,.
\ee
and
\be
\label{eqn:boundcond} {e^{2 i \delta_\ell(i\kappa) }}-1
 =i^{2\nu-1}\pi\frac{I_\nu (\kappa a)}{K_\nu (\kappa a)}
\ee
Inserting Eq.\ (\ref{eqn:boundcond}) into Eq.\ (\ref{eqn:gennec}),
we find for general $m$,
\bea
\label{eqn:hardgennec} T_{\mu\nu}V^\mu V^\nu
= &&\frac{4N_m(r)}{\pi^2}\sum_\ell \int_0^\infty d\kappa\,
\frac{I_\nu (\kappa a)}{K_\nu (\kappa a)}\kappa^2 \left[ \left(
D^m_\ell -C^m_\ell \frac{v_\varphi^2}{\kappa^2 r^2}\right)
K_\nu(\kappa r)^2\right.\\
& &\hspace{2.2in}\left.-D^m_\ell v_r^2 \left(\frac{\ell}{\kappa
r}K_\nu(\kappa r)-K_{\nu+1}(\kappa r)\right)^2\right]\nonumber
\eea and for $m = 3$, \bea \label{eqn:hardm3gennec}
T_{\mu\nu}V^\mu V^\nu &=& \frac{1}{4\pi^2 r}\sum_\ell (2\ell+1)
\int_0^\infty d\kappa\, \frac{I_{\nu} (\kappa a)}{K_{\nu} (\kappa
a)}\kappa^2 \left[ \left(1 -\frac{\ell (\ell+1)
v_\varphi^2}{2\kappa^2
r^2}\right) K_{\nu}(\kappa r)^2 \right. \nonumber \\
& &\hspace{2.2in}\left. - v_r^2 \left(\frac{\ell}{\kappa
r}K_{\nu}(\kappa r)-K_{\nu+1}(\kappa r)\right)^2\right]\,.
\eea

\subsection{NEC numeric results 3+1 dimensions}

\begin{figure}
\begin{center}
\leavevmode\epsfbox{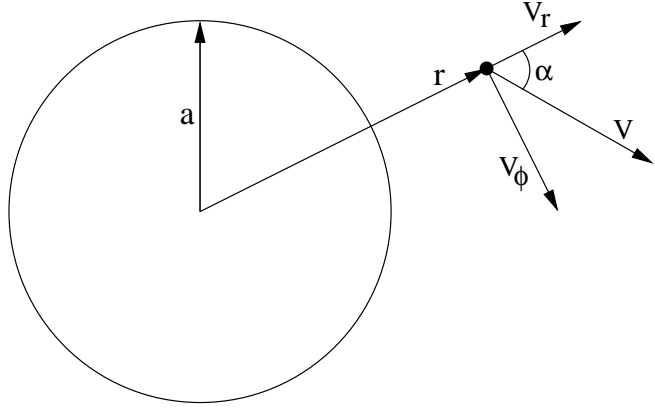}
\end{center}
\caption{We consider the null energy condition at distance $r$
from the center of a sphere for motion in the direction $V$ which
makes an angle $\alpha$ with the radial direction.}
\label{fig:alpha}
\end{figure}

Equation (\ref{eqn:hardm3gennec}) gives $T_{\mu\nu}V^\mu V^\nu $ in
terms of the radial distance from the sphere and the components of
velocity in the radial and azimuthal directions.  We
can rewrite it in terms of the angle
$\alpha$ that any arbitrarily chosen velocity vector makes with the
radial direction, as depicted in Fig. \ref{fig:alpha}, so that
$v_\varphi^2 = \sin^2{\alpha}$, and $v_r^2 = \cos^2{\alpha}$.
We find \bea \label{eqn:hardm3alpha} T_{\mu\nu}V^\mu V^\nu &=&
\frac{1}{4\pi^2 r}\sum_\ell (2\ell+1) \int_0^\infty d\kappa\,
\frac{I_{\nu} (\kappa a)}{K_{\nu} (\kappa a)}\kappa^2 \left[
\left(1 -\frac{\ell (\ell+1) \sin^2{\alpha}}{2\kappa^2
r^2}\right) K_{\nu}(\kappa r)^2 \right.  \\
& &\hspace{2.5in}\left. - \cos^2{\alpha} \left(\frac{\ell}{\kappa
r}K_{\nu}(\kappa r)-K_{\nu+1}(\kappa r)\right)^2\right]\,.\nonumber
\eea

\begin{figure}
\begin{center}
\leavevmode\epsfbox{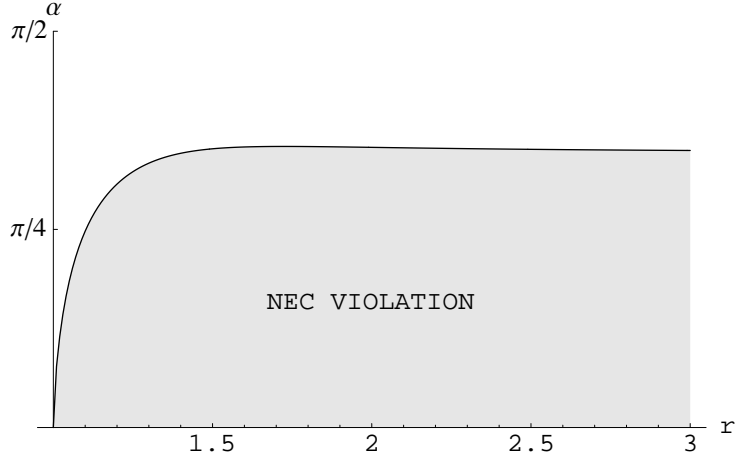}
\end{center}
\caption{Region of NEC violation (shown shaded) in 3+1 dimensions.
The parameter $r$ is the radius in units of $a$, and $\alpha$ is the
angle between $V$ and the radial direction.}
\label{fig:necgraph}
\end{figure}

We have used Mathematica to determine numerically where the NEC is
and is not obeyed.  Figure \ref{fig:necgraph} shows the values for
which the NEC is violated, as a function of radial
distance $r$ and angle $\alpha$.  Radial vectors always violate the
NEC, while azimuthal vectors always obey it.  The dividing line
depends on the distance from the sphere.

\subsection{ANEC in 3+1 dimensions}
\label{sec:anec}

\begin{figure}
\begin{center}
\leavevmode\epsfbox{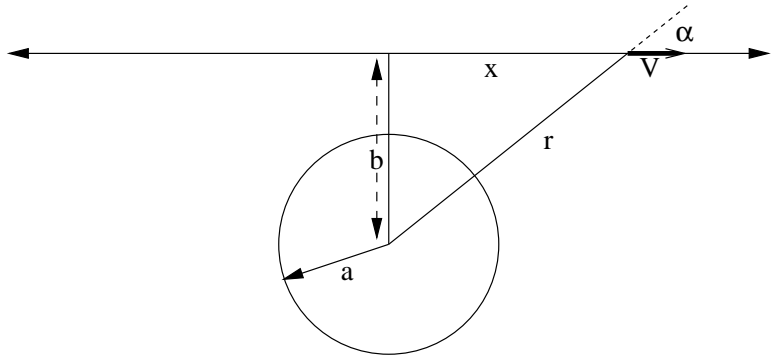}
\end{center}
\caption{We integrate the NEC along a line parallel to the $x$-axis
with impact parameter $b$.}
\label{fig:geodesic}
\end{figure}

We obtain the ANEC by integrating the NEC along a complete null
geodesic.  We consider an observer moving along a geodesic parallel to
the $x$-axis and passing by the spherical boundary with an impact
parameter $b$, as depicted in figure \ref{fig:geodesic}.
We find
\bea
\int_{-\infty}^\infty dx T_{\mu\nu}V^\mu V^\nu &=&
\frac{1}{2\pi^2} \sum_\ell (2\ell+1) \int_{0}^\infty dx
\int_0^\infty d\kappa\, \left(\frac{I_{\nu}(\kappa
a)}{K_{\nu}(\kappa a)}\right)\kappa^2 \frac{1}{r}
\\ && \times \left[ \left(1 -\frac{\ell (\ell+1) v_\varphi^2}{2\kappa^2
r^2}\right) K_{\nu}(\kappa r)^2 - v_r^2 \left(\frac{\ell}{\kappa
r}K_{\nu}(\kappa r)-K_{\nu+1}(\kappa r)\right)^2\right]\nonumber
\eea
where $r^2=x^2+b^2$, $v_\varphi^2 = \sin^2\alpha = b^2/(x^2 + b^2)$,
and $v_r^2 = \cos^2 \alpha = x^2/(x^2 + b^2) $.

\begin{figure}
\begin{center}
\leavevmode\epsfbox{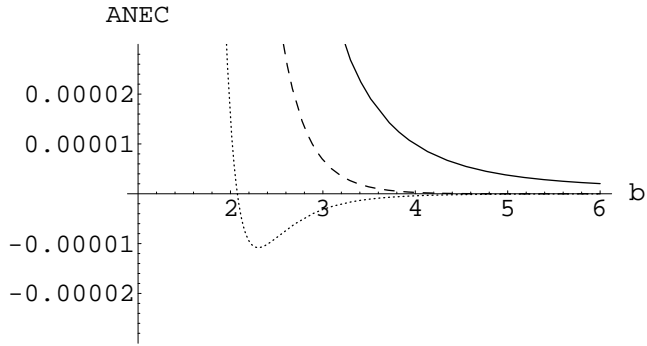}
\end{center}
\caption{Contributions to the ANEC integral from the first three
partial waves ($l = 0$ solid, $l = 1$ dashed and $l = 2$ dotted) in
3+1 dimensions vs impact parameter $b$, with $a = 1$.}
\label{fig:partialwaveanec}
\end{figure}

\begin{figure}
\begin{center}
\leavevmode\epsfbox{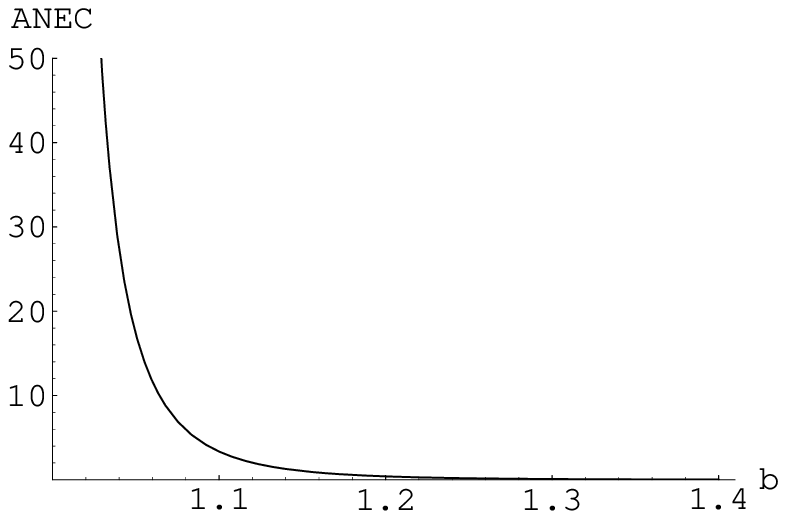}
\end{center}
\caption{ANEC integral in 3+1 dimensions vs. impact parameter $b$, with $a = 1$.}
\label{fig:anec}
\end{figure}

We have found numerically that the ANEC is indeed obeyed.  The sum
over partial waves converges rapidly.  The $\ell=0$ partial wave
always obeys the NEC, but the other partial waves do have negative
contributions for certain impact parameters b, as shown in Fig.\
\ref{fig:partialwaveanec}.  However these negative contributions
are always more than compensated for by positive contributions
from lower $\ell$, as can be seen in Fig.\ \ref{fig:anec}, which
shows the contribution including all partial waves.

\section{NEC and ANEC in 2+1 Dimensions}
\label{sec:2d}

The results obtained for 2+1 dimensions closely follow those for
3+1 dimensions.  For $m = 2$, we find $\nu = l$ and $N_m (r) =
1/8$.  With $l=0$, we use $D_0^2 = 1$, $C_0^2 = 0$ and Eq.\
(\ref{eqn:gennec}) becomes \be \label{eqn:m2l0gennec}
T_{\mu\nu}V^\mu V^\nu = \frac{i}{2\pi^3}\int_0^\infty d\kappa\,
\left(e^{2i \delta_\ell(i\kappa)}-1\right)\kappa^2 \left[
K_0(\kappa r)^2- v_r^2 K_1(\kappa r)^2\right]\,. \ee For $l > 0$,
we use $D_\ell^2 = 2$, $C_\ell^2 = 2l^2$ and we get \bea
T_{\mu\nu}V^\mu V^\nu &=& \frac{i}{\pi^3} \sum_\ell (-1)^\ell
\int_0^\infty d\kappa\, \left(e^{2i
\delta_\ell(i\kappa)}-1\right)\kappa^2 \\ &&
\left[\left(1-\frac{\ell^2v_\varphi^2}{\kappa^2 r^2}\right)
K_\ell(\kappa r)^2-v_r^2 \left(\frac{\ell}{\kappa
r}K_{\ell}(\kappa r)-K_{\ell+1}(\kappa
r)\right)^2\right]\,.\nonumber \eea

With the perfectly reflecting circular boundary condition for
$l=0$ we obtain \be \label{eqn:hardm2l0gennec} T_{\mu\nu}V^\mu
V^\nu = \frac{1}{2\pi^2}\int_0^\infty d\kappa\, \frac{I_0 (\kappa
a)}{K_0 (\kappa a)}\kappa^2 \left[ K_0(\kappa r)^2- v_r^2
K_1(\kappa r)^2\right] \ee and for $l > 0$ we find \bea
T_{\mu\nu}V^\mu V^\nu &=& \frac{1}{\pi^2} \sum_\ell \int_0^\infty
d\kappa\, \frac{I_\ell (\kappa a)}{K_\ell (\kappa a)}\kappa^2 \\
&& \left[\left(1-\frac{\ell^2v_\varphi^2}{\kappa^2 r^2}\right)
K_\ell(\kappa r)^2-v_r^2 \left(\frac{\ell}{\kappa
r}K_{\ell}(\kappa r)-K_{\ell+1}(\kappa
r)\right)^2\right]\,.\nonumber \eea

\begin{figure}
\begin{center}
\leavevmode\epsfbox{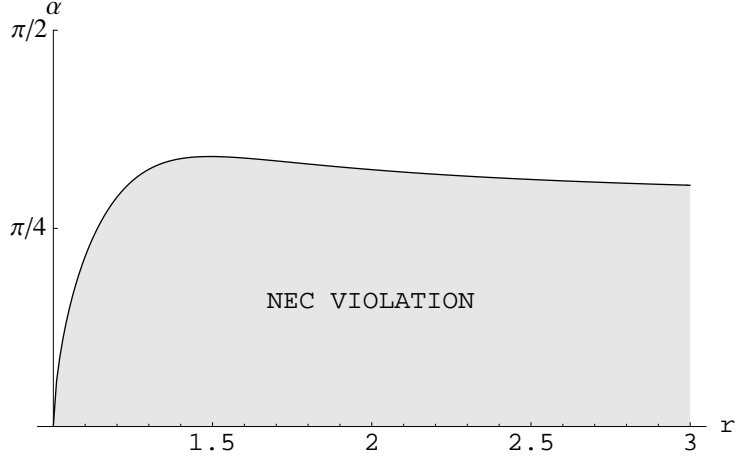}
\end{center}
\caption{Region of NEC violation (shown shaded) in 2+1 dimensions.}
\label{fig:discnecgraph}
\end{figure}

Figure \ref{fig:discnecgraph} shows the values for which the 2+1
dimensional NEC is violated, as a function of radial distance $r$ and
angle $\alpha$.  Once again, radial motion violates the NEC.

\begin{figure}
\begin{center}
\leavevmode\epsfbox{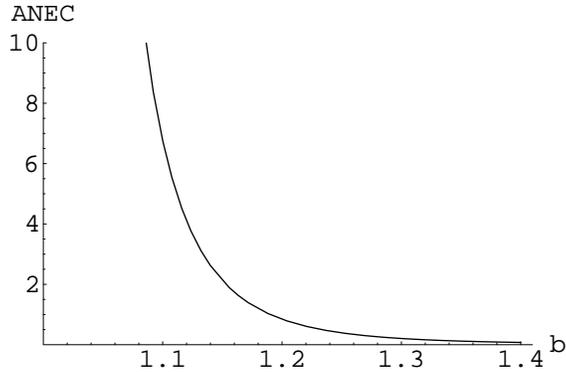}
\end{center}
\caption{ANEC integral in 2+1 dimensions vs. impact parameter $b$,
with $a = 1$.}
\label{fig:anecdisc}
\end{figure}

As in 3+1 dimensions, the ANEC is always obeyed, as shown in
Fig.\ \ref{fig:anecdisc}.

\section{Conclusions}

We have studied the problem of a minimally coupled, real, massless
scalar field outside a spherically symmetric background potential,
within the calculational framework developed in \cite{OlumGraham}.  We
obtained a general expression for the null energy condition in $d$
dimensions.  We calculated the NEC for the specific case of a
perfectly reflecting sphere and showed that the NEC is obeyed at some
points in space and violated at others, depending on the velocity of
the observer.

We calculated the averaged null energy condition for 3+1 and 2+1
dimensions by integrating the NEC over a complete null geodesic.
Although there is a range of impact parameters for which partial waves
with $\ell>0$ contribute negatively to the ANEC, the ANEC is always
obeyed when summing over all the partial waves.

Although we have done explicit calculations only for perfectly
reflecting boundary conditions, we conjecture that the ANEC is
satisfied for all geodesics which pass outside any localized background
potential.

\section{Acknowledgments}

We would like to thank Xavier Siemens for suggesting the problem, Noah
Graham for valuable assistance, and Larry Ford, Tom Roman, and
Alexander Vilenkin for helpful conversations. K. D. O. was supported
in part by the National Science Foundation.

\appendix
\section{Analytic properties of the phase shift}
In this appendix we show that, for positive, real $k$,
\be\label{eqn:Agoal}
(-k)\left(e^{2i \delta_\ell(-k)}-1\right)H^{(1)}_\nu(-kr)^2
= k\left(e^{-2i \delta_\ell(k)}-1\right)H^{(2)}_\nu(kr)^2\,
\ee
where $-k$ is taken in the upper half plane,
by using the techniques of \cite{Graham:2002xq}.  From
Eqs.\ (22) and (21) of \cite{Graham:2002xq}, and using the notation of
that paper, we find the Green's function,
\be
G_l (r, r, k) =\psi_l (k, r) f_l (k, r)\frac{e^{-\pi i (\nu-1/2)}}{k}\,.
\ee
Outside the potential, the Jost solution is just the free Jost solution,
\be
f_l (k, r) = e^{\pi i \nu}\sqrt{\frac{\pi kr}{2}} H^{(1)}(kr)
\ee
while the physical scattering solution can be written in terms of the
scattering phase shift,
\be
\psi_l (k, r)=\sqrt{\frac{\pi kr}{2}}
\left[{e^{2 i \delta_\ell(k)}}H^{(1)}_\nu(kr) + H^{(2)}_\nu(kr)\right]\,.
\ee
Thus
\bea
G_l (r, r, k)&=&\frac{i\pi r}{2} H^{(1)}_\nu(kr)\left[{e^{2 i
\delta_\ell(k)}}H^{(1)}_\nu(kr) + H^{(2)}_\nu(kr)\right]\\
&=&\frac{i\pi r}{2}\left[\left ({e^{2 i
\delta_\ell(k)}}-1\right)H^{(1)}_\nu(kr)^2
+2J_\nu(kr)H^{(1)}_\nu(kr)\right]\,.\nonumber \eea

Now $G$ has the property $G (r, r,-k) = G (r, r,k)^*$, and thus we
find \be \left({e^{2 i
\delta_\ell(-k)}}-1\right)H^{(1)}_\nu(-kr)^2
+2J_\nu(-kr)H^{(1)}_\nu(-kr) =-\left({e^{-2 i
\delta_\ell(k)}}-1\right)H^{(2)}_\nu(kr)^2
 -2J_\nu(kr)H^{(2)}_\nu(kr)\,.
\ee
The second terms on the two sides are equal, so the first terms must
be equal also, which proves Eq.\ (\ref{eqn:Agoal}).

As $k\to -k$, $\tilde H^{(1)}_\nu (kr)^2$ transforms in the same
way as
$H^{(1)}_\nu (kr)^2$, so
\be
(-k)^3\left(e^{2i \delta_\ell(-k)}-1\right)\tilde H^{(1)}_\nu(-kr)^2
= k^3\left(e^{-2i \delta_\ell(k)}-1\right)\tilde H^{(2)}_\nu(kr)^2\,
\ee
as well.

\end{document}